\documentclass[12pt]{article}
\usepackage{amsmath}
\usepackage{graphicx}
\usepackage{natbib}
\usepackage{multirow}
\usepackage{url} 

\newcommand{\blind}{0}

\begin{document}

\def\spacingset#1{\renewcommand{\baselinestretch}%
{#1}\small\normalsize} \spacingset{1}


\if0\blind
{
  \title{\bf Statistical Evolution of ODI Cricket: Analyzing Performance Trends and Effect Sizes}
  \author{Pratik Mullick\hspace{.2cm}\\
    Department of Operations Research and Business Intelligence,\\
    Wrocław University of Science and Technology,\\
    50-372 Wrocław, Poland}
  \maketitle
} \fi

\if1\blind
{
  \bigskip
  \bigskip
  \bigskip
  \begin{center}
    {\LARGE\bf Title}
\end{center}
  \medskip
} \fi

\bigskip
\begin{abstract}
In the dynamic realm of One Day International (ODI) cricket, the sport has undergone significant transformations over the past four decades. This study digs into the intricate evolution of ODI cricket from 1987 to 2023, analyzing about 4000 matches to uncover pivotal performance indicators such as batting prowess, bowling efficiency, and partnership dynamics. Employing statistical methodologies, including Cohen's effect size, we scrutinize the observed changes that have shaped ODI cricket's landscape. Our findings reveal nuanced trends: while first innings scores have shown stability with sporadic high outliers in recent years, the impact of achieving scores exceeding 300 has notably increased. Furthermore, batting depth and early wickets lost in the first innings continue to significantly influence match outcomes, highlighting strategic shifts in team approaches. We also observe improvements in second innings bowling effectiveness, particularly in wicket-taking ability, underscoring evolving defensive strategies. This research contributes a statistical foundation to comprehensively understand the evolving dynamics of ODI cricket, offering insights crucial for strategic decision-making and further analysis in sports analytics.
\end{abstract}

\noindent%
\vfill

\newpage
\spacingset{1.75} 
\section{Introduction}
\label{sec:intro}

Cricket, a sport that has captured the hearts of millions worldwide, stands as a testament to the enduring spirit of competition and unity. Originating in medieval England, cricket has evolved into one of the most loved and widely followed sports, extending its influence far beyond its birthplace. Unlike many sports, cricket boasts a unique blend of strategy, skill, and endurance, captivating audiences with its dynamic and often unpredictable nature. With its roots dating back to the 16th century, cricket has not only weathered the test of time but has flourished into a global phenomenon, played and celebrated across continents.

At its essence, cricket involves two teams taking turns to bat and bowl, with the objective of scoring more runs than their opponents. The game unfolds on a large oval field, known as the pitch, where a bowler from the fielding team attempts to dismiss the batsmen from the opposing team, known as getting a wicket. Meanwhile, the batters strive to score runs by hitting the ball and running between two sets of stumps, separated by 22 yards. Cricket matches can span various formats, from the leisurely-paced Test matches that unfold over several days to the high-octane Twenty20 games that condense the excitement into a few hours. With its rich history, diverse formats, and the thrill of last-minute turnarounds, cricket has entrenched itself as more than just a sport -- it's a cultural phenomenon that unites people in their love for the game.

Statistical analysis has always been an integral part of any game, as demonstrated by numerous studies in sports such as football \citep{football,football2}, basketball \citep{basketball,basketball2}, and baseball \citep{baseball0,baseball}. In cricket, study of match statistics offers an empirical foundation for understanding player performances, team strategies, and match outcomes. Through the meticulous collection and analysis of data \citep{cricket1,cricket2,cricket3,cricket4}, statisticians can identify patterns and trends that might not be immediately apparent, providing insights that are vital for both on-field decisions and long-term planning. Statistical metrics such as batting averages, strike rates, and bowling economies help quantify player contributions, while advanced techniques like regression analysis and effect size calculations reveal deeper dynamics of the game. By transforming raw data into actionable insights, statistics enhance the strategic depth of cricket, guiding coaches, players, and analysts in optimizing performance and crafting winning strategies.

Over the years, cricket has undergone a remarkable transformation, with numerous changes and adaptations contributing to the dynamic nature of the sport. One of the most significant shifts occurred with the introduction of limited-overs formats, such as One Day Internationals (ODIs) and Twenty20 (T20) matches, injecting a dose of excitement and urgency into the game. The traditional Test format, known for its leisurely pace, witnessed alterations in rules to accommodate day-night matches, enhancing the spectator experience. Technological advancements, like the implementation of Decision Review System (DRS), have brought a new layer of precision to umpiring decisions, sparking debates and adding an intriguing dimension to the game.

Furthermore, innovations like Powerplays in ODIs and strategic timeouts in T20s have revolutionized team strategies, providing captains with tactical tools to navigate the ebb and flow of the game. The concept of player auctions in T20 leagues has transformed the cricketing landscape, fostering a global community of players and fans alike. From the introduction of white cricket balls for day-night matches to the tweaks in fielding restrictions, each modification has left an indelible mark on cricket's ever-evolving narrative. As cricket continues to adapt to the demands of the modern era, these changes underscore the sport's resilience and its ability to captivate audiences through innovation and reinvention.

In the dynamic realm of modern cricket, the study of performance analysis \citep{2007wc,wicketloss,womenst20,Tharoor2022PerformanceOI,Doljin2015DevelopmentOA,cricket_perform1} has become increasingly vital, reflecting the sport's multifaceted evolution. Analyzing various performance indicators \citep{PI} provides a thorough understanding of player strategies, team dynamics, and the intricate balance between bat and ball. Investigating how these indicators influence the outcome of matches not only deepens our appreciation for the game but also offers valuable insights for players, coaches, and cricket enthusiasts. With the introduction of limited-overs formats and strategic innovations, the significance of performance metrics has soared, guiding teams in crafting winning strategies and enhancing player performances .

The thorough analysis of batting averages, bowling figures, and partnership statistics unveils patterns and trends that shape the narrative of a match. As cricket enthusiasts, researchers, and analysts delve into the world of performance analysis, they uncover the impact of rule changes, technological advancements, and shifting player dynamics on the game's competitive landscape \citep{bagladesh,choosingbatsmen,TummaSusmitha2023AnalysisAP}. The study of performance indicators, often quantified through metrics like Cohen's effect size, acts as a compass navigating through the complexities of cricket, shedding light on the factors that tip the scales in favor of victory or defeat. These analyses are also useful to study individual player performances, that are essential for team management to pick players from an auction\citep{choosingbatsmen,MUKHERJEE2014624}. In essence, performance analysis in cricket not only decodes the nuances of the sport but also lays the foundation for strategic advancements, fostering a continuous dialogue that propels the game forward.

In this paper, we conduct a thorough examination of the dynamic landscape of One Day International (ODI) cricket. The inception of ODI cricket dates back to January 5, 1971 \citep{bateman_cambridge}. The ODI Cricket World Cup tournaments held in 1975, 1979, and 1983, referred to as Prudential Cups at the time, featured matches with 60 overs per innings. The transition to the current standard of 50 overs per innings in ODI matches occurred in the 1987 edition of the cricket world cup, starting on October 8. This 50-overs-per-innings format remains the prevailing norm in ODI cricket. To ensure the uniformity of our analysis and interpretation, we exclusively considered data from October 8, 1987, through the conclusion of the 2023 world cup on November 19, 2023. Our aim is to analyse the intricate web of performance indicators that shape the outcome of ODI matches and understand how these indicators have transformed over the years. With a collection of extensive data on batting, bowling, and partnership dynamics, we perform rigorous statistical analysis to quantify the impact of these factors on match results.

It is a common perception that contemporary cricket heavily favors batsmen, leading to frequent discussions about the diminishing role of bowlers and the imbalance in favor of batting. However, this study aims to challenge this notion by conducting a comprehensive analysis of key performance indicators over the past few decades. Specifically, we seek to investigate how both batting and bowling performances have influenced match outcomes in ODI cricket. By examining first innings scores, run rates, and wickets taken, particularly in the second innings, our goal is to provide a nuanced understanding of the evolving dynamics of the game and to determine whether bowling still plays a decisive role in securing victories.

The investigation involves analysis of performance metrics, employing statistical tools such as Cohen's effect size $d$ to gauge the magnitude of differences in various aspects of the game. As we traverse through different eras, we seek to unravel the trends, patterns, and shifts in performance that have shaped ODI cricket. By examining the statistical details, we provide an insight of how the game has evolved and how key performance indicators have influenced success or failure in the fast-paced and unpredictable world of ODIs.

This paper contributes to the broader discourse on cricket research by combining historical insights with rigorous statistical analysis. Our findings not only offer a retrospective glance at the evolution of the sport but also shed light on the contemporary landscape of ODI cricket. Ultimately, this endeavor aims to enrich the appreciation for the game, offering valuable insights for players, coaches, and enthusiasts keen on understanding the dynamics of cricket's past and present.

\section{Materials and Methods}
\label{sec:meth}

\subsection{Data}

\begin{figure}[h!]
    \centering
    \includegraphics[width=\textwidth]{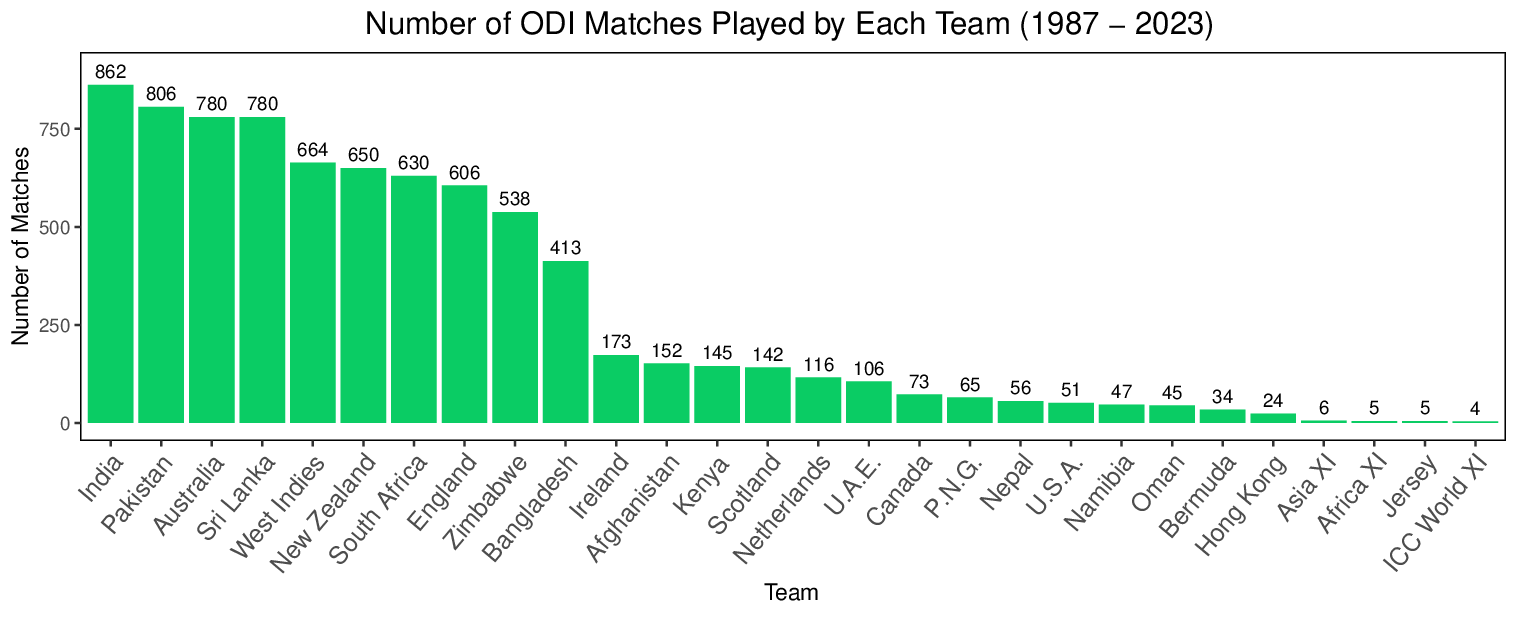}
    \caption{Number of ODI cricket matches played by 28 different teams within the span of October 8, 1987 to November 19, 2023. The data shown here consists of only those matches which resulted in a winning team, i.e., excludes the matches with results `Tied' and `N/R'. }
    \label{fig_team_matches}
\end{figure}

The raw data used in this research were collected from the ESPN cricinfo's online repository called STATSGURU\footnote{link: \url{https://stats.espncricinfo.com/ci/engine/stats/index.html}}. From October 8, 1987 to November 19, 2023 we found a total of 4255 ODI matches recorded in the database. These matches were played among 28 different teams at 179 different venues over the globe. Out of the 28 teams, 25 were independent ODI-playing nations and there were 3 composite teams, viz. ICC World XI, Asia XI and Africa XI, which were formed for specific matches or series, representing players from the respective regions rather than individual countries. To begin our analysis, we discarded the matches which did not result in a winner, i.e., `Tied' and `No Result' (abandoned) matches. This left us with a record of 3989 ODI matches. Figure \ref{fig_team_matches} summarises the number of matches played by each of the 28 teams in our dataset.  The team batting data consisted of the total scores, overs batted, wickets fallen in each innings and the year of the match. For records of partnership between two batters in a match we collected the data for number of runs scored in each wicket of the innings. We found a total of 64658 partnerships in our chosen time frame of analysis presented in this article.

We organized the extensive dataset spanning the years 1987 to 2023 into distinct temporal segments to facilitate a comprehensive analysis of the evolution of ODI cricket over time. Specifically, the data was divided into 8 year-groups, as summarised in Table \ref{tab1}. This segmentation allowed us to capture the dynamic changes and trends in ODI cricket performance across these 8 distinct periods. The data presented in Table \ref{tab1} underscores the ample sample sizes at our disposal, ensuring robust statistical analyses with substantial statistical power.
\begin{table}[h!]
\caption{Distribution of ODI macthes across 8 year-groups within the span 1987 to 2023. The data with asterisks (*) are too small in size to be considered for analysis of effect size, and they were discarded.}
{\begin{tabular}{lcccccccc}
\hline
 Year-group & Matches &\multicolumn{3}{c}{Matches with first innings score} & \multicolumn{4}{c}{Partnerships} \\
 & & $200 - 249$ & $250 - 299$ & $\geq 300$ & Openning & Top & Middle & Lower\\
 \hline
 $1987 - 1990$ & 205 & 81 & 46 & 7* & 412 & 1228 & 1370 & 572 \\
 $1991 - 1995$ & 345 & 132 & 81 & 12* & 705 & 2101 & 2342 & 976 \\
 $1996 - 2000$ & 606 & 209 & 161 & 57 & 1222 & 3631 & 4131 & 1885 \\
 $2001 - 2005$ & 609 & 170 & 191 & 82 & 1244 & 3685 & 4135 & 1894 \\
 $2006 - 2010$ & 712 & 202 & 184 & 130 & 1479 & 4387 & 4970 & 2387 \\
 $2011 - 2015$ & 587 & 157 & 177 & 129 & 1218 & 3608 & 4132 & 1971 \\
 $2016 - 2020$ & 511 & 126 & 135 & 145 & 1048 & 3113 & 3556 & 1691 \\
 $2021 - 2023$ & 414 & 98 & 119 & 96 & 837 & 2495 & 2901 & 1494 \\
 \hline
\end{tabular}}
\label{tab1}
\end{table}

\subsection{Performance Indicators}\label{pi}

Performance indicators (PI) in the context of cricket are specific metrics or statistics used to evaluate the performance of players and teams during a match \citep{PI}. These indicators help quantify various aspects of the game, providing insights into strengths, weaknesses, and overall effectiveness. In Table \ref{tab2} we summarize the PI's that have been explored in this contribution.
\begin{table}[h!]
\caption{List of performance indicators used in this research. The indicators with an asterisk (*) are the ones which are not normally distributed. Indicators with a dagger ($\dagger$) were studied separately for both the innings.}
{\begin{tabular}{ll}
\hline
 Performance Indicators & Definition\\
 \hline
 First innings score & Total number of runs scored when batting first\\
 First innings score $200-249$ & Total number of runs scored between $200$ and $249$\\ & when batting first\\
 First innings score $250-299$ & Total number of runs scored between $250$ and $299$\\ & when batting first\\
 First innings score $\geq 300$ & Total number of runs scored greater than or equal to $300$\\ & when batting first\\
 Overs batted in first innings* & Total number of overs played when batting first\\
 Run rate of first innings & The average runs scored per over when batting first\\
 Wickets lost in first innings* & Total number wickets lost when batting first\\
 Runs conceded in second innings & Total number of runs conceded when bowling second\\
 Wickets taken in second innings* & Total number of wickets taken when bowling second\\
 Opening partnership*$\dagger$ & Total number of runs scored before the fall of $1^{\text{st}}$ wicket\\
 Top order runs*$\dagger$ & Total number of runs scored before the fall of $3^{\text{rd}}$ wicket\\
 Middle order runs*$\dagger$ & Total number of runs scored between the $3^{\text{rd}}$ and $7^{\text{th}}$ wicket\\
 Lower order runs*$\dagger$ & Total number of runs scored after the fall of $7^{\text{th}}$ wicket\\
 \hline
\end{tabular}}
\label{tab2}
\end{table}
In Section \ref{sec:results}, we have provided detailed analysis of how these PI's have reshaped the game of cricket over the span of 37 years. In the following we briefly describe each of these performance indicators, and their importance in the game.

First innings score is a critical performance indicator in ODI cricket, as it sets the foundation for the rest of the match. The first innings score refers to the total number of runs scored by the team that bats first. This score is pivotal because it dictates the target that the second batting team needs to chase. A high first innings score puts pressure on the opposing team, often forcing them into a more aggressive and risk-laden approach. Conversely, a low first innings score can shift momentum in favor of the chasing team, allowing them to play more conservatively. The ability to post a competitive first innings score is indicative of a strong batting lineup and strategic acumen, and it significantly influences the overall outcome of the match.

In the context of first innings score as a performance indicator, we decided to consider specific score ranges that can significantly impact the outcome of a match. (A) First innings score between 200 - 249: this score range can be considered a competitive yet vulnerable target in ODI cricket. Historically, teams setting scores between 200 and 249 often find themselves in closely contested matches. (B) First innings score between 250 - 299: scores in this range are generally regarded as strong targets in ODI cricket. A first innings score between 250 and 299 usually puts pressure on the chasing team, requiring them to maintain a steady scoring rate while managing the risk of losing wickets. (C) First innings score above 300: achieving a first innings score above 300 has become more common with the evolution of aggressive batting strategies and improved batting techniques. Such scores typically indicate a dominant batting performance and are often associated with high winning probabilities. In Table \ref{tab1} the number of matches relevant for the above mentioned PI's have been recorded. The groups which were too small in size were discarded for the analysis of effect size.

The number of overs batted in the first innings and the run rate of the first innings are vital performance indicators in ODI cricket. Batting through the full 50 overs demonstrates a team's ability to sustain their innings and maximize scoring opportunities, often leading to competitive totals. The run rate, calculated as the average number of runs scored per over, reflects the team's batting aggression and efficiency. Higher run rates typically result in more formidable targets, placing pressure on the opposition.

The number of wickets lost in the first innings is another crucial performance indicator in ODI cricket. It reflects a team's batting stability and resilience under pressure. Losing fewer wickets generally allows a team to maintain momentum, build partnerships, and achieve higher scores. Conversely, frequent wicket losses can disrupt the batting order, restrict scoring, and limit the team's ability to set substantial targets.

Runs conceded and wickets taken in the second innings are critical performance indicators for assessing a team's bowling effectiveness and defensive capabilities in ODI cricket. The runs conceded measure the bowlers' ability to contain the opposition's scoring rate and defend the target set in the first innings. Meanwhile, the wickets taken indicate the bowlers' success in breaking partnerships and applying pressure on the batting side. Together, these indicators provide a comprehensive view of a team's defensive strategy and effectiveness in controlling the game during the crucial second innings.

Partnerships and runs scored at different batting positions are also key PI's that reflect a team's batting strength and stability in both innings. The opening partnership sets the foundation for the first innings, providing a steady start and building momentum, and plays a crucial role in chasing totals in the second innings. Top order runs, scored by the top three or four batsmen, are essential as they anchor the innings and set the pace in the first innings, or stabilize the chase in the second innings. Middle order runs are vital for maintaining the innings' flow and recovering from early losses, often steering the team through the middle overs in both innings. Lower order runs, scored by the tail-end batsmen, can be decisive in adding valuable runs towards the end of the first innings, turning a competitive score into a winning total, or in clinching close chases in the second innings.

\subsection{Statistical Analysis}\label{stat.ana}

The objective of this investigation is to examine the evolution of performance in One Day International (ODI) cricket throughout the years. To achieve this, we identify various performance indicators in the realm of cricket and evaluate their impact on the outcome of ODI matches—whether they result in victory or defeat. The estimation is conducted using Cohen's effect size $d$ \citep{cohen1988}, a widely adopted statistic in the field of performance analysis in cricket and sports \citep{batter} in general. The effect size $d$ signifies the absolute differences in each performance indicator, denoted here by $p$, providing a measure of the magnitude of distinction between winning ($W$) and losing ($L$) teams. Through this analysis, we gain insights into the relative importance of diverse indicators, elucidating their contributions to a successful match outcome. Cohen's $d$ is defined as the difference between two means divided by a standard deviation for the data, i.e., \begin{equation}
    d=\frac{\Bar{p}_W-\Bar{p}_L}{s},
\end{equation} where $s$ is the pooled standard deviation of the data for two independent samples, defined as: \begin{equation}
    s=\sqrt{\frac{(n_W-1)s_W^2+(n_L-1)s_L^2}{n_W+n_L-2}},
\end{equation} and the variance for each of the groups is defined as \begin{equation}
    s_X^2=\frac{1}{n_X-1}\sum_{i=1}^{n_X}(p_{X,i}-\Bar{p}_X)^2,\hspace{0.3cm}X=W\text{ or, }L.
\end{equation} The subscripts $W$ and $L$ denotes the group of data for winning and losing teams respectively. So, $p_W$ (or $p_L$) denotes the set of values of the indicator $p$ for the winning (or losing) team, with $\Bar{p}$ being the mean value. The size of each group is denoted by $n$. The criteria \citep{cohen1988,batter,sawilosky} for interpreting effect size $d$ are summarised in Table \ref{tab3}.\begin{table}[h!]
\caption{Classification of Effect Sizes in Cohen's Scale. This table categorizes Cohen's effect sizes ($d$) into distinct ranges to provide a qualitative interpretation of their magnitudes. The classification spans from Very Small to Huge, with corresponding numerical intervals indicating the extent of the effect size.}
{\begin{tabular}{c|cccccc}
\hline
Effect size & Very small & Small & Medium & Large & Very Large & Huge\\
\hline
$d$&$<0.2$&$0.2-0.5$&$0.5-0.8$&$0.8-1.2$&$1.2-2$&$>2$\\
\hline
\end{tabular}}
\label{tab3}
\end{table}

In our study, we use Cohen's effect size ($d$) to quantify the impact of various factors on cricket match outcomes. The calculation of $d$ involves statistical inference, where the accuracy of the effect size estimation depends on the size of the data sample. Larger samples generally yield more accurate results compared to smaller ones. We apply the concept of confidence intervals (CIs) to gauge the potential error associated with the effect size. For instance, a 90\% confidence interval means there is a 10\% chance of a type I error in estimating the effect size. Interpreting the CI for effect size follows a similar rationale to CIs for means. In our context, if the 90\% CI includes 0, it suggests statistical non-significance. Reporting both the point estimate of the effect size and its CI is crucial for comprehending the magnitude and precision of the observed effects. Assuming that the data is normally distributed, the formula for estimating CI for the effect size $d$ is given by \citep{HEDGES198575}: \begin{equation}
    \sigma(d)=\sqrt{\frac{n_W+n_L}{n_W\times n_L}+\frac{d^2}{2(n_W+n_L)}}
\end{equation}and \begin{equation}
    \text{90\% CI for Cohen's }d: [d-1.64\sigma(d),d+1.64\sigma(d)]
\end{equation} The value $1.64\sigma(d)$ serves as a representation of the 90\% confidence level (CL) for the effect size $d$. In our findings, we have provided this confidence level alongside each estimated effect size.

\subsection{Bootstrap: To Deal with Non-Normal Data}\label{boots}

Cohen's $d$ is a measure of effect size that quantifies the standardized mean difference between two groups. The methodological details presented in Section \ref{stat.ana} is valid when the underlying data is normally distributed. The data for several performance indicators considered in this research follow normal distribution, with exceptions like number of wickets, number of overs in an innings and partnership runs. For large sample sizes, Cohen's $d$ is robust to violations of normality, however, for small sample sizes this might not be true. Our sample sizes for each group are $\mathcal{O}(100)$, and we decided not to exploit the assumptions of normality while estimating effect size.

In our study, we employed bootstrap resampling as a robust method to estimate Cohen's $d$, especially for non-normally distributed data. Bootstrap resampling involves iterative sampling, with replacement, from the observed data to create multiple datasets. In this paper, the number of bootstrap samples created were kept fixed to $10000$ for partnership data and $1000$ for each of the remaining cases. For each of these datasets, we computed Cohen's $d$, capturing the variability in the effect size estimation. This approach is particularly advantageous for non-normally distributed data, where traditional assumptions of normality may not hold. By repeatedly resampling from the observed data, we obtained a distribution of Cohen's $d$, allowing us to assess its stability and variability. This resampling technique is valuable in scenarios where the underlying data distribution is unknown or deviates from normality, providing a robust means of estimating effect sizes in such conditions.

\section{Results}
\label{sec:results}

In this section we describe the significance of each performance indicator described in Section \ref{pi}, and present the corresponding results of statistical analyses which show how the game of cricket has evolved over the years.

\begin{figure}[h!]
    \centering
    \includegraphics[width=\textwidth]{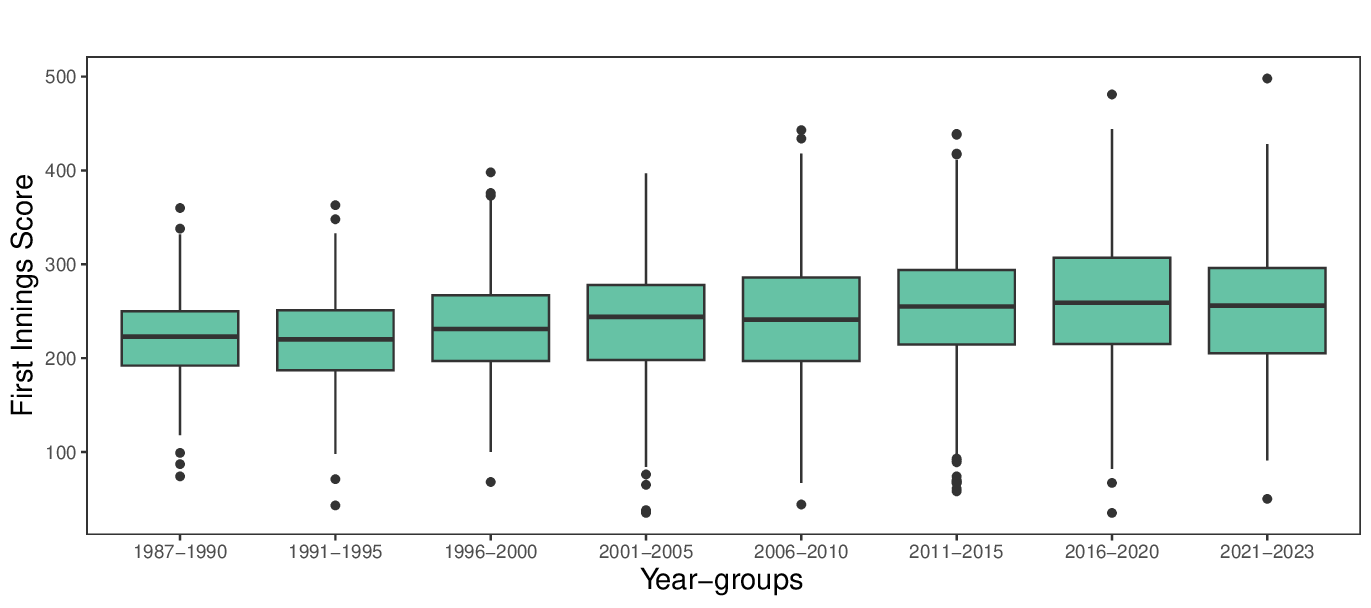}
    \caption{Boxplots for the first innings scores of 8 year-groups that has been considered in this paper. A one-way analysis of variance reveals [$F(7,3981)=23.15$, $p<10^{-6}$, $\eta^2=0.04$] that the scores across these year-groups have small-medium statistical difference. }
    \label{fig_first_innings}
\end{figure}

 In Figure \ref{fig_first_innings} we show the distributions of first innings scores over the years from 1987 to 2023.  This box plot reveals a stable central trend in first innings scores in ODI cricket from 1987 to 2023. However, it also highlights a significant increase in the frequency and magnitude of exceptionally high scores in recent years. This trend could be due to various factors, such as changes in batting techniques, improvements in player fitness, better quality of cricketing equipment, or rule changes favoring batsmen. The rise in high outlier scores suggests that modern ODIs may be witnessing more aggressive batting strategies and higher scoring games, reflecting an evolving dynamic in the sport.

\begin{figure}[h!]
    \centering
    \includegraphics[width=\textwidth]{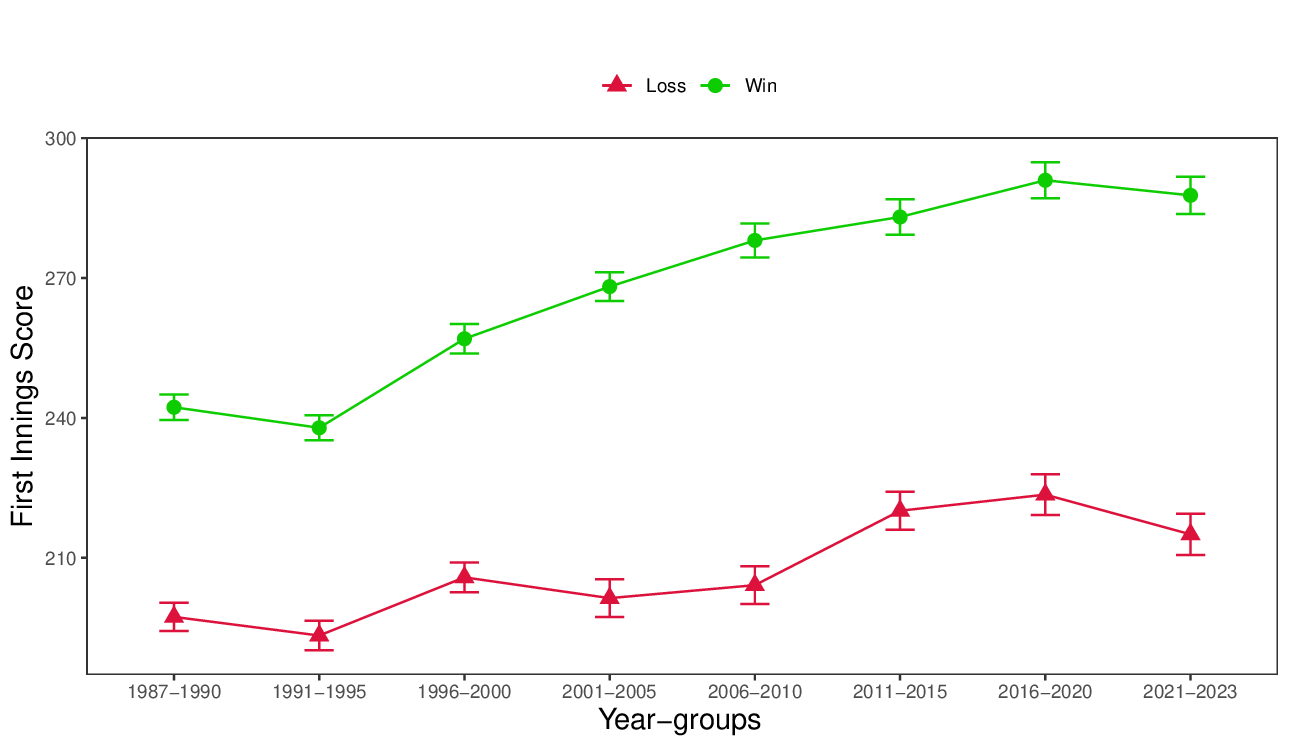}
    \caption{Mean first innings scores across the 8 year-groups shown separately for winning and loosing causes. The error-bars indicate the corresponding standard errors of mean, calculated as standard error = standard deviation/$\sqrt{N}$, where $N$ is the sample size.}
    \label{fig_first_innings_win_loss}
\end{figure}

Higher first innings scores are crucial for increasing the likelihood of winning an ODI match, and this has been a non-trivial consistent pattern across the periods studied, as shown in Figure \ref{fig_first_innings_win_loss}.
 The increase in mean scores over time suggests evolving strategies and performances in ODI cricket. There is a clear trend indicating that higher first innings scores are associated with winning outcomes across all groups. The gap between the winning and losing scores varies, but winning scores consistently exceed losing scores. Over the periods, the mean scores for both winning and losing teams show an increasing trend, reflecting changes in the game dynamics, possibly due to rule changes, improved batting techniques, or other factors. The standard errors (SE) suggest the variability around the mean scores, with slightly higher SEs in the losing scores in some groups, indicating greater inconsistency in those performances.

\begin{figure}[h!]
    \centering
    \includegraphics[width=\textwidth]{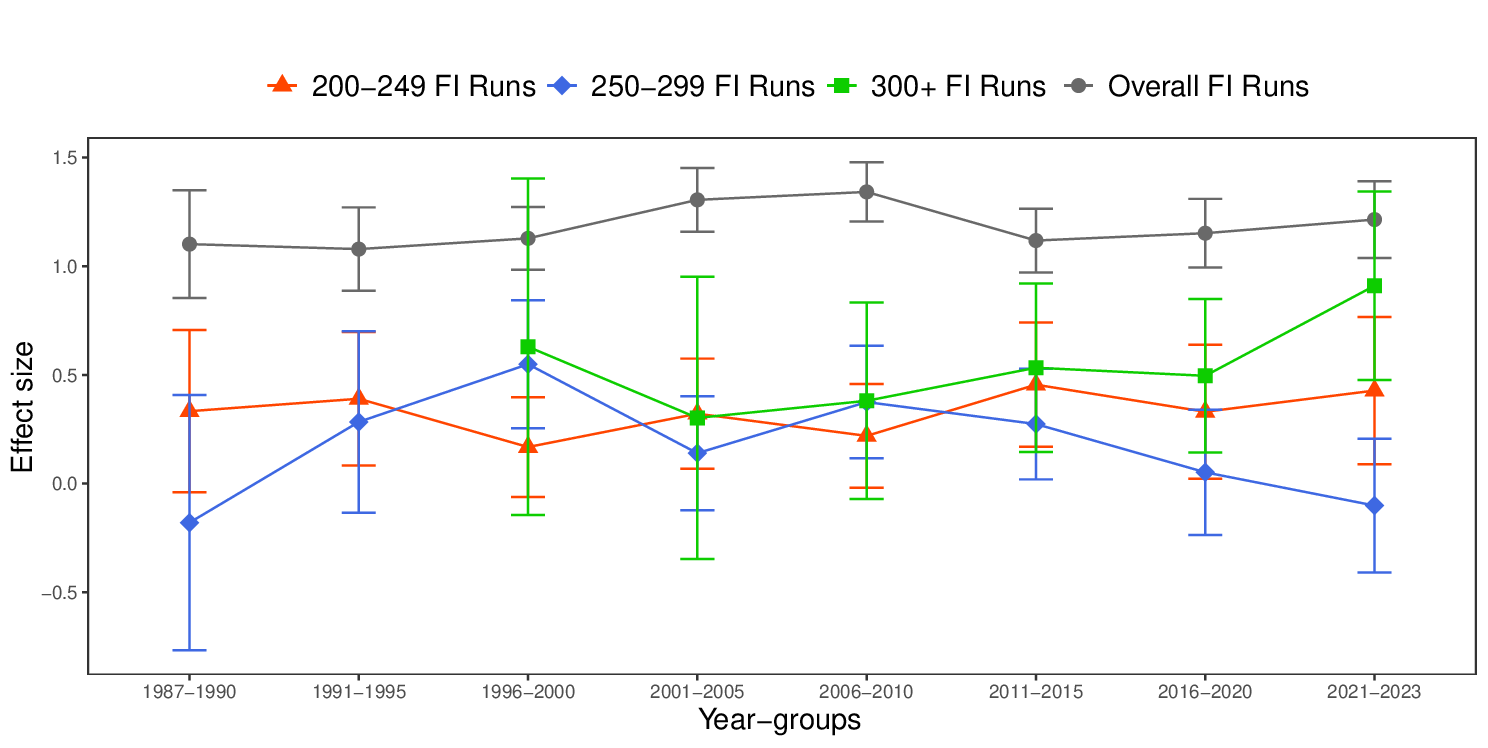}
    \caption{Time dependence of effect sizes (Cohen's d) representing first innings (FI) runs over 8-year groups from 1987 to 2023. This analysis was done for 4 different performance indicators, viz. the overall first innings score and 3 other score ranges mentioned in Table \ref{tab2}. The error bars indicate the $90\%$ confidence intervals.}
    \label{es_FI_runs}
\end{figure}

It was found that the effect sizes observed across different performance indicators related to the first innings score vary in magnitude with time, without showing a regular monotonic behavior, as shown in Figure \ref{es_FI_runs}. 
 The effect size related to the overall first innings score demonstrates a significant impact, ranging from 'large' to 'very large' (all $d>1$), across all matches analyzed. For matches with first innings scores between 200 and 249, and 250 to 299, effect sizes vary within the 'small' to 'medium' range, indicating moderate yet discernible influences on player performances and match dynamics. Notably, matches featuring first innings totals exceeding 300 reveal an increasing trend in effect sizes over time, suggesting a growing significance of high-scoring encounters. The effect size for matches with first innings scores exceeding 300 consistently surpasses those of the two lower score ranges, except for the year-group 2001-2005. This highlights the statistically significant association between achieving a 300+ score in the first innings and winning matches.

It was very interesting to note that first innings scores ranging from 250-299 shows a consistent increase in effect size until 2000, but a decline post-2006. This implies that in the pre-2000 era, achieving a total within this range had a progressively greater impact on match outcomes. This could indicate a trend where teams scoring within this range were more likely to win matches or exert significant pressure on their opponents. However, the impact of achieving a total within 250-299 on match outcomes decreased after 2006. Teams scoring in this range may have experienced less success or faced greater challenges in converting these scores into match-winning performances.

\begin{figure}[h!]
    \centering
    \includegraphics[width=\textwidth]{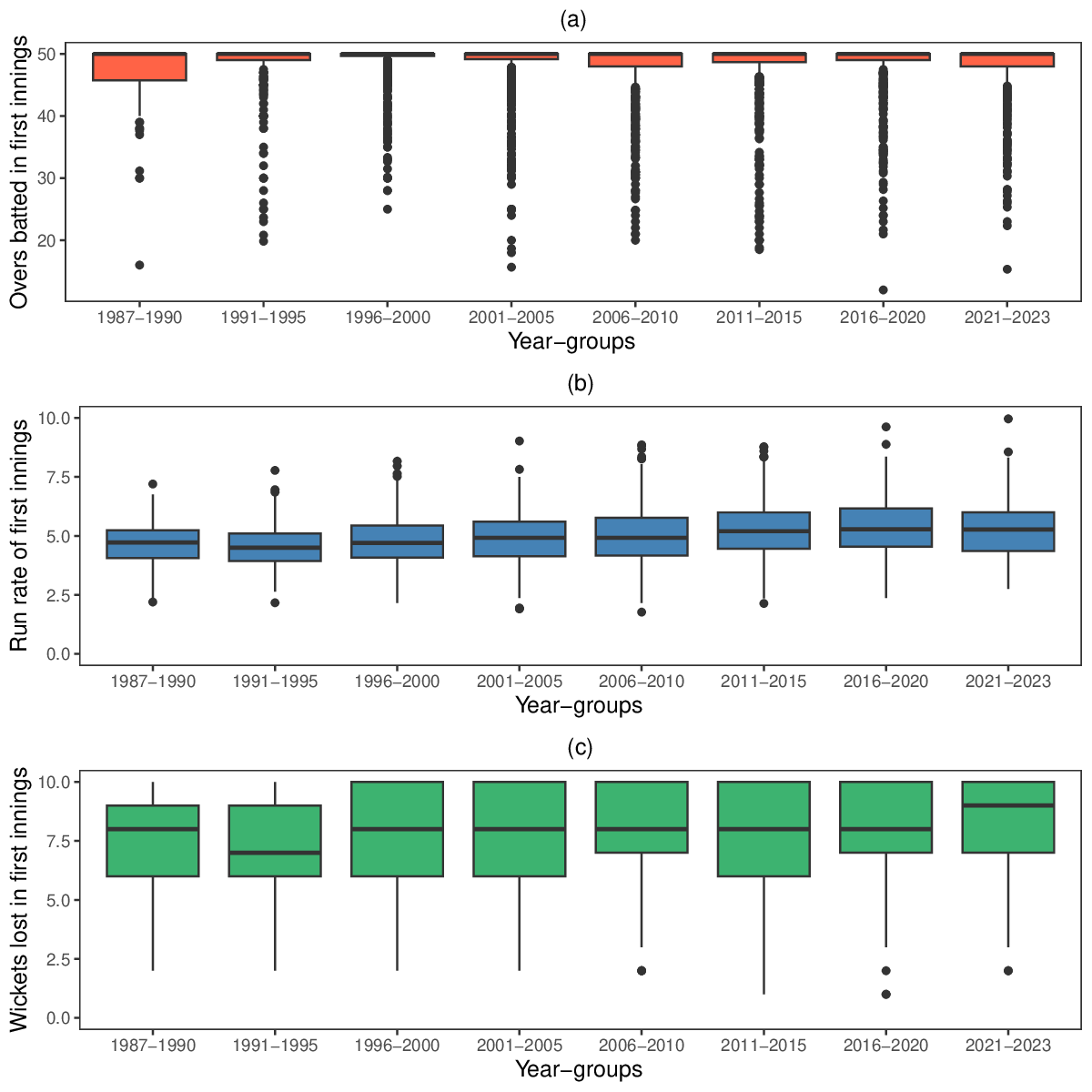}
    \caption{Box plots for (a) overs batted in first innings, (b) run rate of first innings and (c) wickets lost in first innings across the 8 year-groups. The data for (a) and (c) are seen to be heavily skewed. Bootstrap resampling was used for these performance indicators before estimating the effect sizes.}
    \label{des_stat_FI}
\end{figure}

Next, we studied several performance indicators to assess the batting performance of team batting in the first innings. These indicators are number of overs, run rate and wickets loss, whose descriptive statistics across the year-groups are shown in Figure \ref{des_stat_FI}.
 The time dependence of the first innings run rate exhibits a pattern similar to that of the first innings score (Figure \ref{fig_first_innings}), which is expected since the run rate is calculated by dividing the first innings score by the number of overs played. However, the data for overs batted and wickets lost show a longer left tail, indicating left skewness and a violation of the normality assumption required for estimating Cohen's effect size. Therefore, using bootstrap resampling enhances the robustness and reliability of our estimations, as mentioned in Section \ref{boots}.

\begin{figure}[h!]
    \centering
    \includegraphics[width=\textwidth]{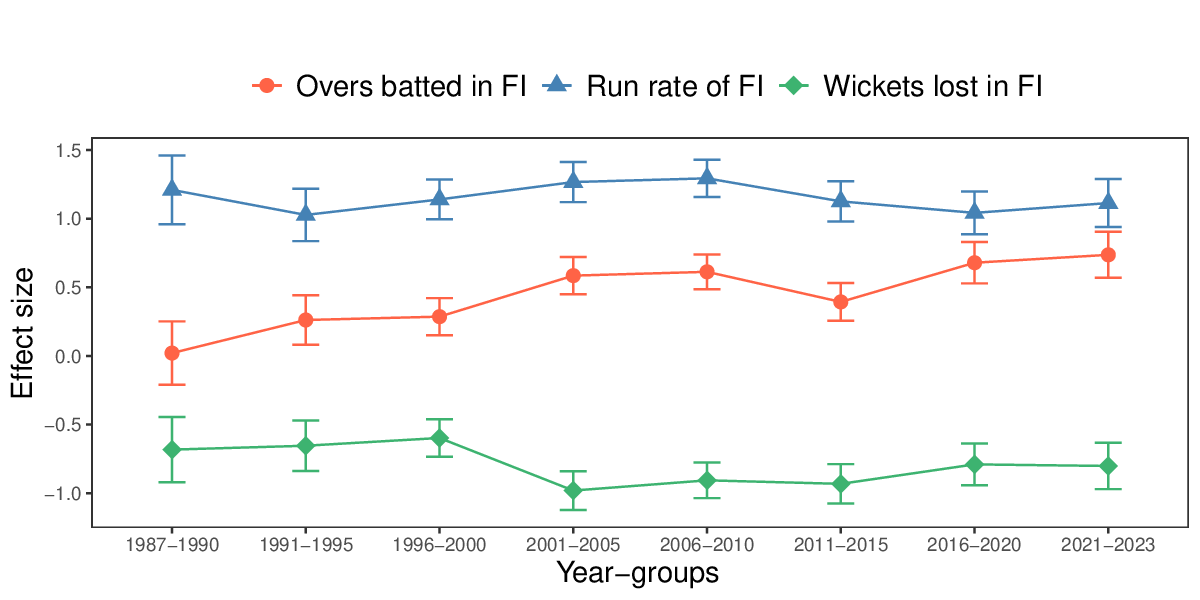}
    \caption{Time dependence of effect sizes (Cohen's d) representing overs batted in first innings (FI), run rate of first innings (FI) and wickets lost in first innings over 8-year groups from 1987 to 2023. The error bars indicate the $90\%$ confidence intervals.}
    \label{es_FI_overs_rr}
\end{figure}

The analysis of effect sizes for the performance indicators overs batted in first innings and run rate of first innings indicates that both the variables have a positive association with match outcomes in ODI cricket, as shown in Figure \ref{es_FI_overs_rr}.
 The effect sizes for overs batted in the first innings demonstrate that from the earlier year groups to the later ones, there is a noticeable increase in the effect sizes, indicating a growing importance of batting through a greater number of overs in the first innings. This trend suggests a shift in batting strategies over the years, with teams increasingly prioritizing building a solid foundation and capitalizing on the opportunities provided by batting throughout a larger proportion of the innings. On the contrary, the effect sizes for the run rate in the first innings exhibit less variability across different year groups. While the overall trend indicates a positive association between run rate and match outcomes, the magnitude of this association remains more or less the same over time.

The comparatively higher effect sizes observed for run rate in the first innings suggest that scoring runs at a brisk pace early in the innings has a more pronounced impact on match outcomes compared to simply batting through a greater number of overs. The disparity in effect sizes highlights the evolving strategic priorities within ODI cricket. Teams are increasingly recognizing the importance of setting a strong tempo early in the innings through aggressive batting and maintaining a healthy run rate. While batting depth and occupying the crease remain important, the emphasis on run rate underscores a shift towards proactive and aggressive batting strategies aimed at dominating the opposition from the outset.

The effect sizes for the number of wickets lost in the first innings are negative across all year groups, as shown in Figure \ref{es_FI_overs_rr}. This negative association suggests that losing wickets early in the innings has a detrimental effect on match outcomes, with a greater number of wickets lost correlating with a decreased likelihood of achieving a favorable outcome in the match. Interestingly, the magnitude of the effect size shows an incremental shift after the 2000s. This suggest that losing wickets early in an ODI match has become more crucial in determining the match's outcome. Teams are paying more attention to the number of wickets they lose early on because it has a bigger impact on their chances of winning. This means they are focusing more on building partnerships and avoiding early collapses to give themselves a better chance of winning. Advanced technology and data analysis tools could perhaps be helping teams understand the significance of early wickets better. Teams these days may be using data to inform their strategies and emphasize the importance of wicket preservation.

\begin{figure}[h!]
    \centering
    \includegraphics[width=\textwidth]{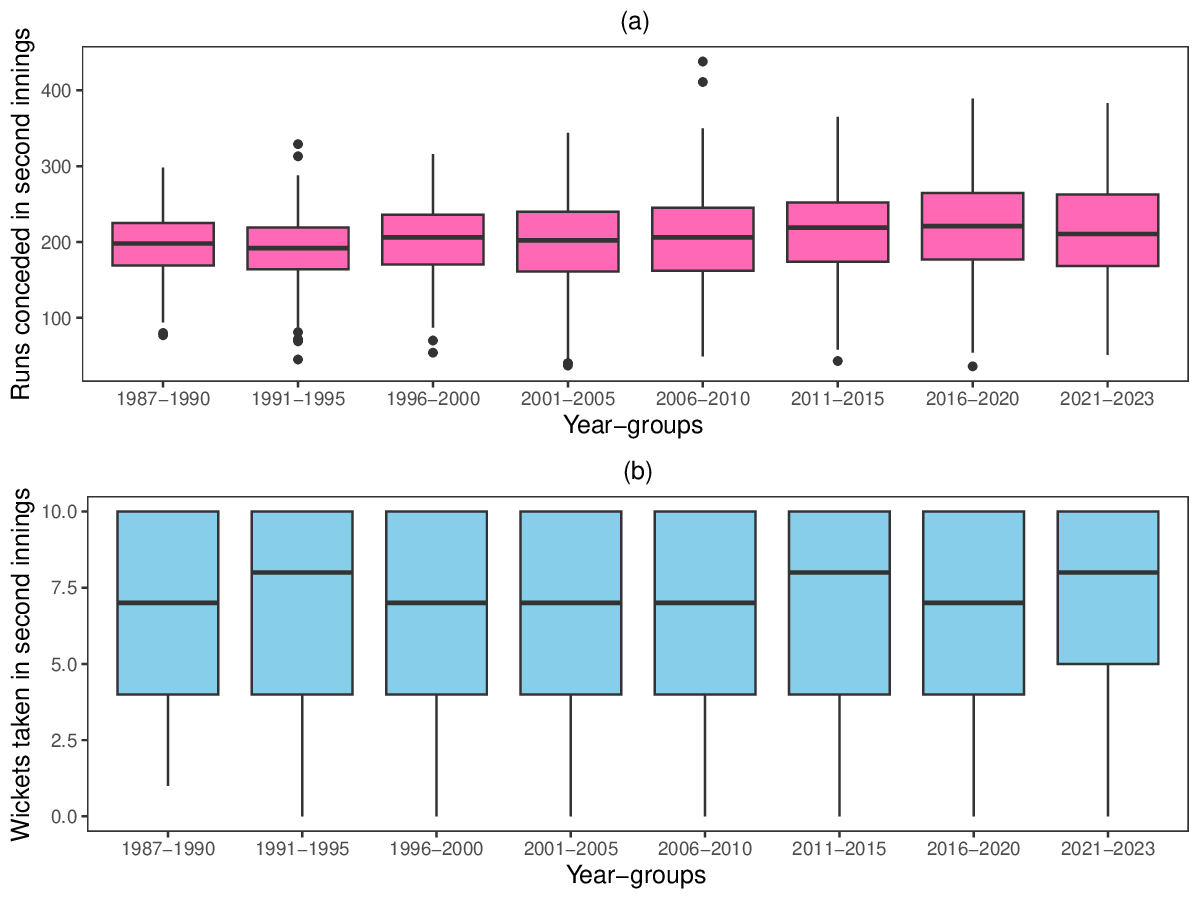}
    \caption{Box plots for (a) runs conceded in second innings and (b) wickets taken in second innings across the 8 year-groups. The data for (b) are seen to be heavily skewed. Bootstrap resampling was used for this performance indicator before estimating the effect sizes.}
    \label{des_stat_SI}
\end{figure}

The performance indicators that were analysed to study the bowling performance of the teams were runs conceded and wickets taken in the second innings. The descriptive statistics of these two indicators over the last 37 years has been shown in Figure \ref{des_stat_SI}.
 The data for runs conceded displays symmetric distributions across all year groups, while the data for wickets taken is left-skewed. This further highlights the need for bootstrap resampling to obtain reliable effect size estimates. Additionally, runs conceded in the second innings exhibit a slightly increasing trend, consistent with the pattern observed in first innings runs scored.

\begin{figure}[h!]
    \centering
    \includegraphics[width=\textwidth]{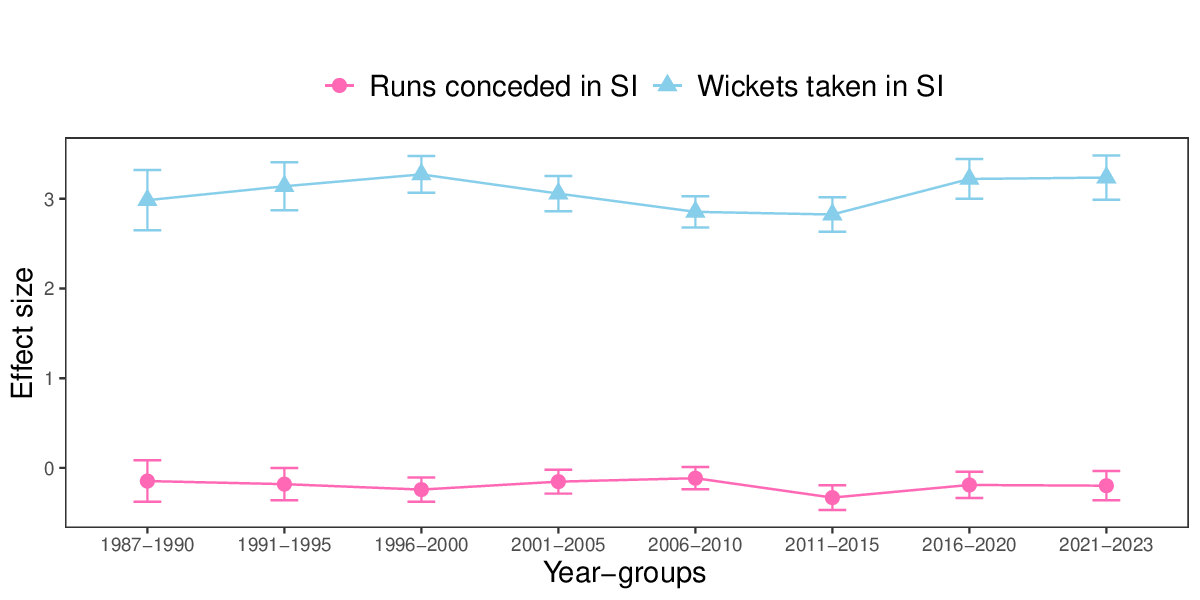}
    \caption{Time dependence of effect sizes (Cohen's d) representing runs conceded in second innings (SI) and wickets taken in second innings over 8-year groups from 1987 to 2023. The error bars indicate the $90\%$ confidence intervals.}
    \label{es_SI_runs_wick}
\end{figure}

The effect size for runs conceded in the second innings shows negative values across all year groups, as depicted in Figure \ref{es_SI_runs_wick}.  This consistent negative effect size indicates that, on average, teams have been conceding fewer runs in the second innings relative to a baseline, suggesting a possible improvement in bowling efficiency or defensive strategies over the years. The confidence intervals for these effect sizes indicate a moderate degree of variability but consistently including zero, which suggests the changes are not statistically significant. In contrast, the effect size for wickets taken in the second innings is remarkably high, with values ranging from $2.824$ to $3.272$ (Figure \ref{es_SI_runs_wick}). These large effect sizes indicate a very strong impact, showing that teams have been significantly more successful in taking wickets during the second innings over the years. The confidence intervals for these effect sizes, ranging from $0.175$ to $0.335$, confirm the statistical significance of this trend, as they do not include zero.

Considering the bowling efficiency of teams in the second innings with regard to the performance indicators described in Figure \ref{es_SI_runs_wick}, we can see a clear and substantial improvement in the ability of teams to take wickets in the second innings over the years, reflected by the high and statistically significant effect sizes. This suggests that bowling strategies or conditions in the second innings have evolved to favor the bowling side significantly. On the other hand, while the runs conceded in the second innings show a trend towards reduction, the effect sizes are small and the confidence intervals suggest that these changes are not statistically significant. This implies that the impact of runs conceded in the second innings has remained relatively stable over time, with no strong evidence of substantial change.

\begin{figure}[h!]
    \centering
    \includegraphics[width=\textwidth]{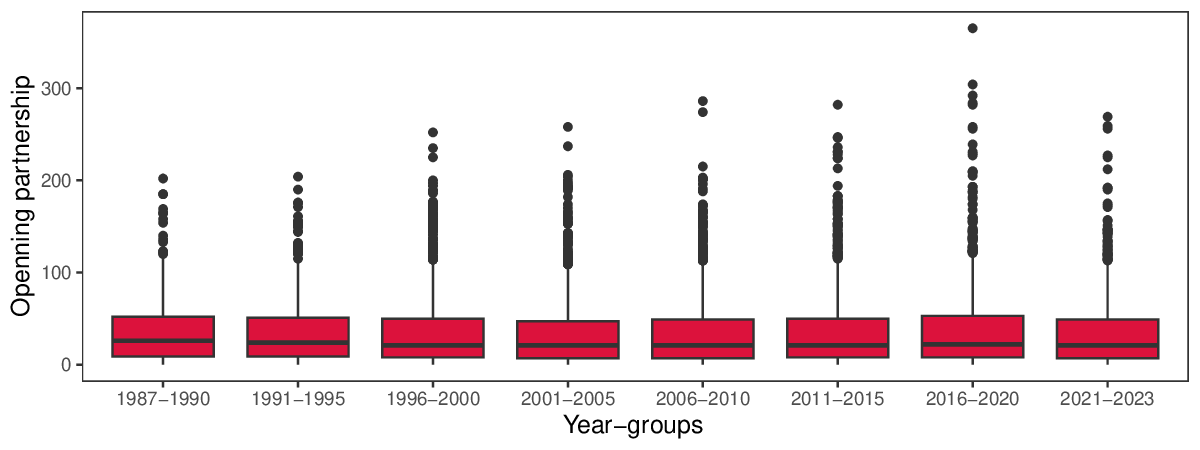}
    \caption{Box plots for opening partnership runs across the 8 year-groups. The data are seen to be heavily skewed. Bootstrap resampling was used for this performance indicator before estimating the effect sizes.}
    \label{des_stat_part_open}
\end{figure}

Finally we studied the partnership data from 1987 to 2023 to study their effect on winning a game. In Figure \ref{des_stat_part_open} we show the descriptive statistics for the first wicket partnership or opening partnership across the 8 year-groups considered in our analysis.  The data does not follow a normal distribution, they have longer right tails and several outliers. Similar skewed distributions are observed when we studied other performance indicators related to partnership data, viz. top order runs, middle order runs and lower order runs. For all these cases bootstrap sampling was used before estimating the effect sizes and their confidence intervals. Overall for the partnership data, the median values for each partnership category show stability with negligible improvements over the years. This suggests consistent performance  in batting across different order positions. The number of high outliers, indicates that batsmen have increasingly had the capability to play exceptional innings. This trend is more prominent in recent years.

\begin{figure}[h!]
    \centering
    \includegraphics[width=\textwidth]{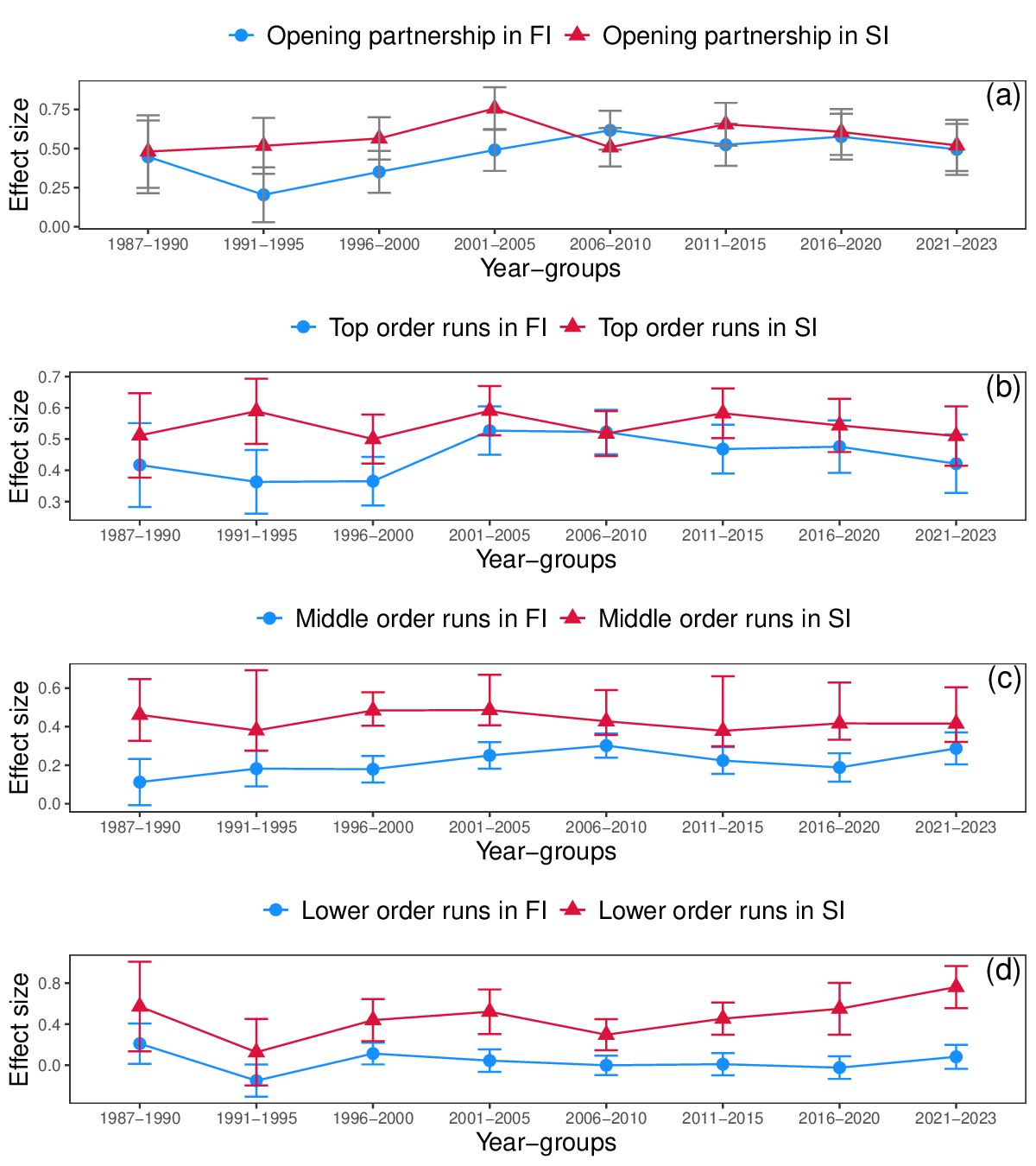}
    \caption{Time dependence of effect sizes (Cohen's d) representing different performance indicators related to partnership runs from 1987 to 2023. These indicators are (a) opening partnership (b) top order runs (c) middle order runs and (d) lower order runs, and they were studied separately for first innings (FI) and second innings (SI).}
    \label{es_part}
\end{figure}

The effect sizes for opening partnerships in both innings demonstrate a consistent pattern over the years, as shown in Figure \ref{es_part}(a). The effect size is generally higher for the second innings (SI) compared to the first innings (FI), indicating that a strong opening partnership is more crucial for chasing scores.  The relatively narrow confidence intervals suggest a stable influence across the years. Similarly, the effect sizes for top order runs exhibit consistent behavior over the years (Figure \ref{es_part}(b)), with wider confidence intervals in the earlier years reflecting variability in top-order contributions. Higher effect sizes in the second innings imply that top-order performance significantly impacts outcomes when chasing a target. The steady effect sizes for middle order runs in recent years (Figure \ref{es_part}(c)) highlight the sustained significance of middle-order performance, with higher effect sizes in the second innings compared to the first. For lower order runs in the second innings (Figure \ref{es_part}(d)), there is a steady increase after 2006, whereas in the first innings, the effect sizes are not statistically significant (or negligible in some cases) as they include zero within the $90\%$ confidence intervals.

The most significant discovery of this analysis is the consistent higher effect sizes in the second innings (SI) compared to the first innings (FI) across all partnership performance indicators. This highlights the critical importance of these partnerships in successfully chasing a target. This pattern is evident in opening partnerships, top order runs, middle order runs, and lower order runs, suggesting that strong performances from battings partnerships at various positions are more crucial when pursuing a score rather than setting one. The higher effect sizes in the SI imply that successful run chases rely heavily on the ability of batting partnerships to withstand pressure and maintain momentum, thereby significantly influencing match outcomes. This trend underscores the strategic emphasis teams place on building and sustaining effective partnerships during the chase, as these contributions often determine the final result of the game.

\section{Discussion}
\label{sec:dis}

In this contribution, we have presented statistical analysis to provide a comprehensive overview of how key performance indicators (PIs) in ODI cricket have evolved and influenced match outcomes from 1987 to 2023. The results offer significant insights into the changing dynamics of the game, reflecting broader trends in strategy, performance, and the overall evolution of ODI cricket.

The stability in the central trend of first innings scores alongside the increase in exceptionally high scores highlights a shift towards more aggressive batting strategies in modern cricket. This evolution likely stems from a combination of improved batting techniques, enhanced player fitness, better equipment, and rule changes favoring batsmen. The consistent association between higher first innings scores and winning outcomes underlines the strategic emphasis on setting formidable targets, which has remained a crucial aspect of match success across all periods studied. The notable rise in mean scores for both winning and losing teams further indicates that teams have adapted to the higher scoring environment by enhancing their offensive capabilities.

The first innings run rate mirrors the trend observed in first innings scores, reinforcing its role as a critical determinant of match outcomes. The analysis reveals a greater effect size for run rate compared to overs batted, emphasizing the importance of maintaining a high scoring rate early in the innings. This finding suggests a strategic shift towards aggressive batting to establish a commanding position early in the game. Meanwhile, the increasing effect sizes for overs batted in the first innings highlight the growing importance of utilizing the full allotment of overs, suggesting that teams prioritize building substantial innings through deeper batting lineups and better resource management.

The negative effect sizes associated with the number of wickets lost in the first innings underscore the detrimental impact of early wickets on match outcomes. The heightened significance of early wicket preservation post-2000 indicates a tactical shift towards building solid partnerships and minimizing early collapses. This trend may be driven by advanced data analysis and technology, enabling teams to better understand and mitigate the risks associated with losing early wickets.

The consistent negative effect sizes for runs conceded in the second innings suggest an improvement in bowling efficiency or defensive strategies over time. This trend, though not statistically significant, indicates a general enhancement in the ability of teams to contain the opposition during the chase. On the other hand, the high effect sizes for wickets taken in the second innings reflect a substantial improvement in the ability of teams to secure breakthroughs when defending totals. This significant trend highlights the evolving effectiveness of bowling strategies and conditions favoring bowlers in the latter stages of the game.

The analysis of partnership data reveals that strong performances from batting partnerships are crucial for successful run chases, more so than for setting a target. The higher effect sizes observed for partnerships in the second innings across all categories (opening, top-order, middle-order, and lower-order runs) underscore the strategic emphasis on building and sustaining partnerships under pressure. This pattern indicates that effective partnerships are vital in maintaining momentum and withstanding pressure during run chases, thereby significantly influencing match outcomes. The steady increase in effect sizes for lower-order runs in the second innings post-2006 suggests an enhanced capability of lower-order batsmen to contribute significantly in high-pressure situations, reflecting an overall improvement in team depth and resilience.

\section{Conclusion}
\label{sec:conc}

Statistical analysis presented in this study challenges the commonly held belief that modern ODI cricket has overwhelmingly become a batsman's game, with bowlers having little influence on the outcome. While it is true that first innings scores have increased over the years, reflecting more aggressive batting strategies and higher scoring games, our findings indicate that the impact of bowling, particularly in the second innings, remains crucial.

Specifically, our results show that the effect sizes for wickets taken in the second innings are significantly higher than those for runs scored in the first innings. This suggests that despite the emphasis on high scores and aggressive batting, the ability to take wickets in the second innings plays a more decisive role in determining the outcome of a match. Teams that can effectively bowl out their opponents or significantly restrict their scoring in the second innings are more likely to secure victories. This finding underscores the importance of bowling strategies and the skill of bowlers in modern ODI cricket.

Contrary to the popular narrative, our analysis reveals that cricket remains a balanced game where both batting and bowling performances are critical. While batsmen have adapted to new techniques and equipment to score more runs, bowlers have also evolved, developing strategies and skills to counter aggressive batting and exert influence in the crucial second innings. This highlights the dynamic nature of the sport and the continuous evolution of both batting and bowling facets, contributing to the overall competitiveness and unpredictability of ODI cricket. These insights are crucial for teams in formulating their strategies, emphasizing the importance of not only setting high first innings scores but also focusing on effective bowling performances in the second innings to maximize their chances of winning.

\bigskip
\begin{center}
{\large\bf SUPPLEMENTARY MATERIALS}
\end{center}

\begin{description}

\item[ODI data set:] Dataset used for the analysis of key performance indicators in One Day International (ODI) cricket from 1987 to 2023. This dataset includes detailed match information, capturing various performance metrics for both teams involved in each match. The data fields include - $year$: the year in which the match was played, venue: the location where the match was held, $team1$: the name of the first team, $team2$: the name of the second team, $team1\_total$: the total runs scored by the first team, $team1\_wickets$: the number of wickets lost by the first team, $team1\_overs$: the number of overs batted by the first team, $team2\_total$: the total runs scored by the second team, $team2\_wickets$: the number of wickets lost by the second team, $team2\_overs$: the number of overs batted by the second team, $winning\_team$: the identifier for the winning team (1 if $team1$ won, 2 if $team2$ won). (.csv file)

\item[Partnership data set:] Dataset on partnership performances in One Day International (ODI) cricket matches from 1987 to 2023. The data fields include - Wkt: the wicket number at which the partnership took place, Runs: the number of runs scored during the partnership, Inns: the innings number (1 for the first innings, 2 for the second innings), Year: the year in which the match was played, $winning\_team$: the identifier for the winning team (1 if the team won, else 0 for loss). (.csv file)

\item[R Code:] One single code file that has been used to perform the analysis and make the plots. This code takes as input the above mentioned data files, and utilizes various library functions in R.

\end{description}

\bibliographystyle{chicago}
\bibliography{main}
\end{document}